\documentclass[11pt]{article}
\usepackage{moriond,epsfig}

\def\Journal#1#2#3#4{{#1} {\bf #2}, #3 (#4)}

\def\NPB{{\em Nucl. Phys.} B}
\def\PLB{{\em Phys. Lett.}  B}
\def\PRL{\em Phys. Rev. Lett.}
\def\PRD{{\em Phys. Rev.} D}
\def\ZPC{{\em Z. Phys.} C}

\def\be{\begin{equation}}
\def\ee{\end{equation}}
\def\bea{\begin{eqnarray}}
\def\eea{\end{eqnarray}}

\begin{document}
\vspace*{4cm}

\title{ ISOLATED PROMPT PHOTON PRODUCTION \footnote
{Invited paper presented by E. L. Berger at the XXXIInd Rencontres 
de Moriond, {\it QCD and High Energy Hadronic Interactions}, Les Arcs, 
Savoie, France, March 22 - 29, 1997.  Argonne report ANL-HEP-CP-97-32.}}  

\author{ EDMOND L. BERGER$^a$, 
         XIAOFENG GUO$^b$, and JIANWEI QIU$^c$
 }

\address{$^a$High Energy Physics Division, Argonne National Laboratory,\\
Argonne, IL 60439, USA \\
	 $^b$Physics Department, Columbia University,\\
NY, NY 10027, USA \\
         $^c$Department of Physics and Astronomy, Iowa State University,\\
Ames, IA 50011, USA }

\maketitle
\abstracts{We show that the conventionally defined partonic cross section 
for the production of isolated prompt photons is not an infrared safe quantity.
We work out the case of $e^+e^- \rightarrow \gamma + X$ in detail, and we 
discuss implications for hadron reactions such as $p \bar{p} \rightarrow 
\gamma + X$.}

\section{Terminology}
\subsection{Photon Isolation}\label{subsec:1.1}
In $e^+e^-$ and in hadron-hadron reactions at collider energies,
prompt photons are observed and their cross sections are measured only
if the photons are relatively isolated in phase space.  Isolation is required  
to reduce various hadronic backgrounds including those from the 
electromagnetic decay of mesons, e.g., $\pi^o \rightarrow 2\gamma$.
The essence of isolation is that a cone of half-angle $\delta$ is drawn about 
the direction of the photon's momentum, and the isolated cross
section is defined for photons accompanied by less than a specified
amount of hadronic energy in the cone, e.g., $E_h^{\rm cone}\leq
E_{\rm max} = \epsilon_h E_{\gamma}$; $E_{\gamma}$ denotes the energy of the 
photon.  Instead of $\delta$, the Fermilab collider groups use the 
variable $R = \sqrt {(\Delta \phi)^2 + (\Delta \eta)^2}$, 
where $\Delta \eta$ and $\Delta \phi$ denote differences of rapidity and 
azimuthal angle variables.  Theoretical predictions will therefore depend 
upon the additional parameters $\epsilon_h$ and $\delta$ (or $R$).  The 
{\it isolated} cross section is not an {\it inclusive} cross section, and, as 
we discuss below, the usual factorization theorems for inclusive cross 
sections do not apply.  Isolation removes backgrounds, but it also reduces the 
signal.  For example, it reduces the contribution from processes in which the 
photon emerges from the long-distance fragmentation of quarks and 
gluons, themselves produced in short-distance hard collisions.  

\subsection{Conventional Factorization}\label{subsec:1.2}

Much of the predictive power of perturbative QCD derives from 
factorization theorems~\cite{css1}.  {\it Conventional} factorization expresses 
a physical quantity as the convolution of a partonic term  
with a nonperturbative long-distance matrix element.  It requires that the 
partonic term, calculated perturbatively order-by-order in the the strong 
coupling strength $\alpha_s$, have no infrared singularities, and that the 
long-distance matrix element be universal.  Applied to 
the case of $e^+e^- \rightarrow A X$, conventional factorization states 
\begin{equation}
\sigma_{e^+e^-\rightarrow A + X}(Q) 
     = \hat{\sigma}_{e^+e^-\rightarrow c+X}(x,Q/\mu) 
     \otimes D_{c\rightarrow A}(z,\mu) + O\left( {{1}\over{Q}}\right). 
\label{eo}
\end{equation}
The intermediate state $c$ may be a quark, antiquark, gluon, or photon.  The 
symbol 
$\otimes$ denotes a convolution.  Variable $z = p_A/p_c$.  The fragmentation 
function $D_{c\rightarrow A}(z,\mu)$ represents long-distance, small momentum 
scale physics; its dependence on the fragmentation scale $\mu$ is governed by 
the Altarelli-Parisi evolution equations.  In situations in which the 
factorization theorem applies, the partonic hard-part (``the partonic cross 
section") $\hat{\sigma}_{e^+e^-\rightarrow c+X}(x,Q/\mu)$ is infrared safe, 
i.e., finite when all infrared regulators are removed.  

Conventional factorization holds for {\it inclusive} prompt photon production 
$e^+e^-\rightarrow\gamma X$, demonstrated through next-to-leading 
order~\cite{BGQ1}.  However, since the isolated cross section is not an 
inclusive quantity, factorization need not hold and indeed does 
not~\cite{BGQ2}.  Nevertheless, almost all existing calculations of the cross 
section for isolated photon production assume its validity~\cite{theory}.  
Following the standard calculational procedures of perturbative quantum 
chromodynamics (QCD), we show that the partonic hard-part for the isolated 
photon cross section is not infrared safe.  The infrared sensitivity shows up 
first in the next-to-leading order quark-to-photon fragmentation 
contribution~\cite{BGQ2}.  We use the terminology ``breakdown of 
conventional factorization" to describe this result.  

\section{Photons in $e^+e^-$ Annihilation, $e^+e^-\rightarrow\gamma + X$}
Electron-positron reactions offer a relatively clean environment for the study 
of prompt photon production in hadronic final states~\cite{exptal1}.  Since 
there are no complications from initial state hadrons, 
$e^+e^-\rightarrow\gamma + X$ is a good process in which to examine QCD 
predictions in the final state, and the data may be a good source of 
information on quark-to-photon and gluon-to-photon fragmentation 
functions~\cite{early,bgq3,exptal2}.  In turn, these fragmentation functions 
are needed for predictions of photon yields in hadronic collisions.  
Hard photons in $e^+e^-$ processes arise as QED bremsstrahlung from the 
initial beams, radiation that is directed along angles near 
$\theta_{\gamma} = 0\ {\rm and}\ \pi$, and as final state radiation from 
{\it direct} and {\it fragmentation} processes.  The topic of interest in 
this paper is the final state radiation.  It populates all angles, with an 
angular distribution having both transverse, 
$1 + {\rm cos}^2 \theta_{\gamma}$, and longitudinal components~\cite{BGQ1}.  

Cataloging the contributions to $e^+e^-\rightarrow\gamma + X$, we may first 
list the lowest order partonic process: $e^+e^-\rightarrow\ q + \bar{q}$, 
followed by quark or antiquark fragmentation into a photon, 
$q \rightarrow  \gamma X$.  The ${\cal O}(\alpha_{em})$ direct contribution is 
represented by the three-body final state process 
$e^+e^-\rightarrow\ q + \bar{q} + \gamma$.  Separation of the lowest order 
fragmentation and the ${\cal O}(\alpha_{em})$ direct contributions is not 
unique; the fragmentation scale $\mu$ dependence relates the two.  At 
${\cal O}(\alpha_s)$ the contributions of interest come from the 
three-body final state processes $e^+e^-\rightarrow\ q + \bar{q} + g$, 
followed by gluon fragmentation, $g \rightarrow \gamma X$; and from 
$e^+e^-\rightarrow\ q + \bar{q} + g$, followed by quark or antiquark 
fragmentation, $q \rightarrow \gamma X$.  In Fig.~1(a) and 1(b), we illustrate 
the set of Feynman diagrams used in the computation of the contribution from 
$e^+e^-\rightarrow\ q + \bar{q} + g$ (with the implied understanding that the 
final quark or antiquark fragments into a photon).  The set of real gluon 
emission diagrams in 
Fig.~1(a) results in an infrared sensitive contribution, associated with 
the region of phase space in which the emitted gluon momentum becomes soft.  
Likewise, the set of virtual gluon loop diagrams in Fig.~1(b) also results in 
an infrared sensitive contribution.  In the case of {\it inclusive} photon 
production, the infrared singularities cancel once the results from the two 
sets are combined.  However, in the {\it isolated} photon case, the restriction 
on the phase space accessible to the final state gluon in the set of Fig.~1(a) 
leads to an incomplete cancellation between the two sets.  The 
${\cal O}(\alpha_s)$ partonic hard 
part $\hat{\sigma}^{(1)}_{e^+e^-\rightarrow qX}(x,Q/\mu)$ is therefore not 
finite when the infrared regulator is removed.  

\begin{figure}
\hbox{{\hskip 1.0cm}\epsfxsize9.5cm\epsffile{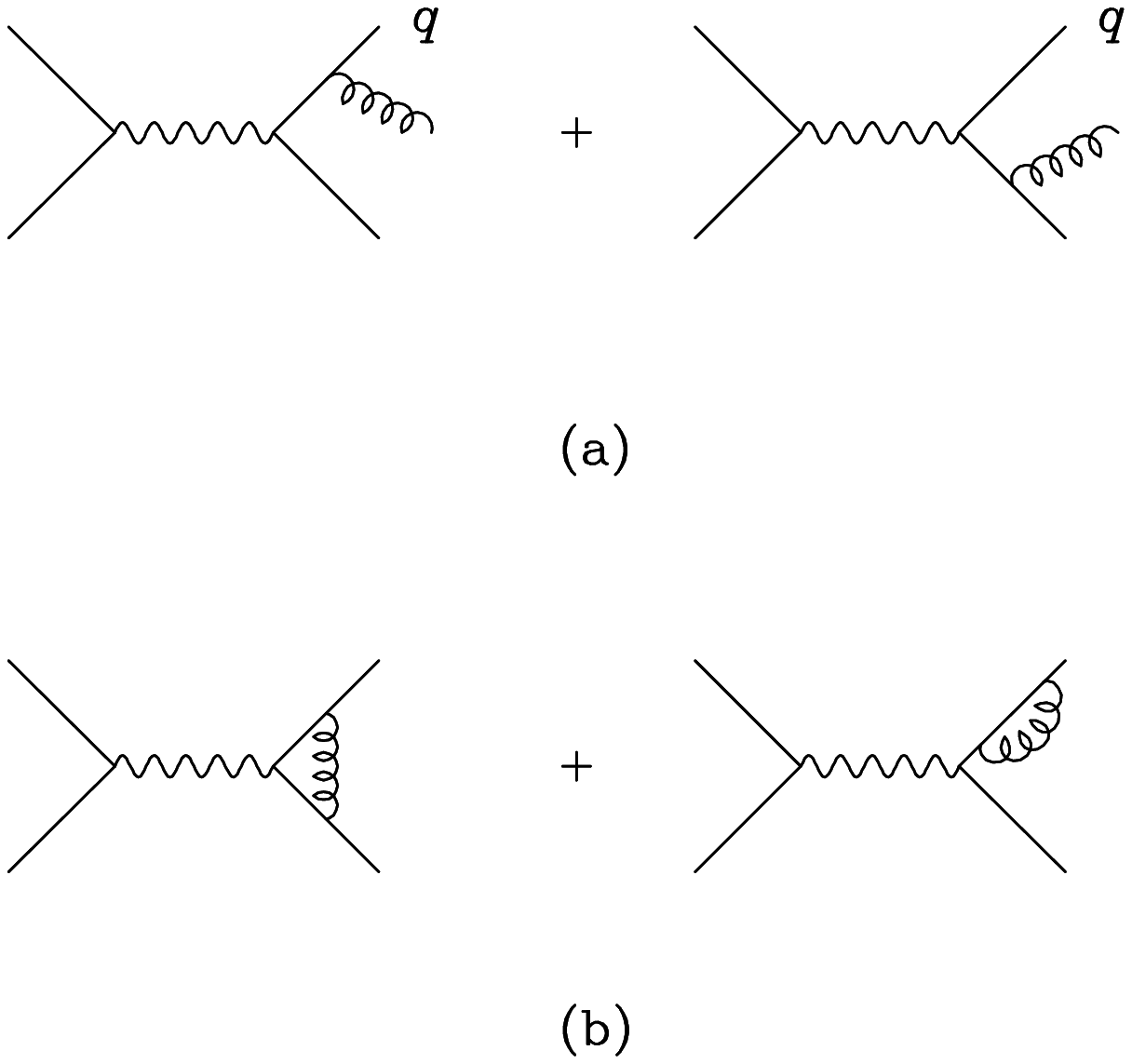}
\epsfxsize9.5cm\epsffile{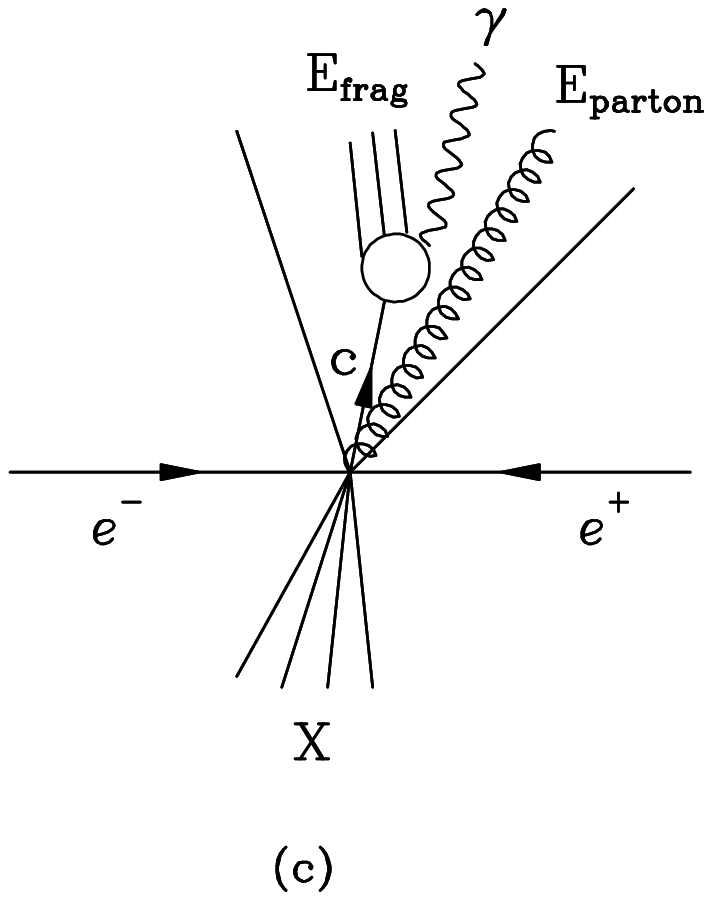}}
\vskip -2.0cm
\caption{Feynman diagrams are shown in (a) and (b) for the $O(\alpha_s)$ cross 
section $\hat{\sigma}^{(1)}_{e^+e^- \rightarrow q X}$; both real gluon emission 
diagrams (a) $(e^+e^- \rightarrow q\bar{q}g)$ and virtual gluon exchange terms 
(b) are drawn.  In (c), an illustration is presented of an isolation cone 
containing a parton $c$ that fragments into a $\gamma$ plus hadronic energy 
$E_{\rm frag}$.  In addition, the cone includes a gluon that fragments giving 
hadronic energy $E_{\rm parton}$.}
\end{figure}

\section{Breakdown of Conventional Factorization}
In this section, we summarize our explicit calculation that demonstrates 
infrared sensitivity of the partonic hard part for isolated prompt photon 
production in $e^+e^- \rightarrow \gamma X$.  It is useful to examine the 
sketch in Fig.~1(c).  In that sketch, we illustrate an isolation cone centered 
on a $\gamma$ produced through fragmentation of a parton $c$.  For the 
fragmentation contributions, there are two sources of hadronic energy in the 
isolation cone: i) $E_{\rm frag}$ from fragmentation of parton $c$ itself, and 
ii) $E_{\rm partons}^{\rm cone}$ from final-state partons other than $c$ 
that also happen to be in the cone.  The total hadronic energy in the cone is 
$E_{\rm hadrons}^{\rm cone} = E_{\rm frag} + E_{\rm partons}^{\rm cone}$.  
For an isolated $\gamma$, $E_{\rm hadrons}^{\rm cone} \le E_{\rm cut}$, where 
$E_{\rm cut}$ denotes the arbitrary limitation on hadronic energy in the cone 
that is selected in experiments.  We choose to write 
$E_{\rm cut} = \epsilon_h E_{\gamma}$, an equation that defines the 
quantity $\epsilon_h$.  When the maximum hadronic energy allowed 
in the isolation cone is saturated by the fragmentation energy, $E_{\rm
cut}=E_{\rm frag}$, there is no allowance for energy in the cone from other
final-state partons.  In particular, if there is a gluon in the final
state, the phase space accessible to this gluon is restricted.  

We make frequent use of the variable $x_{\gamma} = 2E_{\gamma}/\sqrt s$, and 
we define 
\begin{equation}
x_{crit} = \frac{1}{1 + \epsilon_h} .  
\label{ec}
\end{equation}  
There are three cases of interest: $x_{\gamma} < x_{crit}$, 
$x_{\gamma} = x_{crit}$, and $x_{\gamma} > x_{crit}$.  Because of the 
isolation condition, the phase space constraints are different in the three 
regions.  We summarize the physical situation in the separate regions and show 
that the next-to-leading order partonic term for 
quark fragmentation, $E_q d\hat{\sigma}^{iso}_{e^+e^- \rightarrow qX}/d^3p_q$,
is infrared sensitive~\cite{BGQ2} {\it at and below} the point 
\hbox{$x_\gamma = 1/(1+\epsilon_h)$}. 

\begin{figure}
\vskip -2.9cm
\hbox{\epsfxsize9.5cm\epsffile{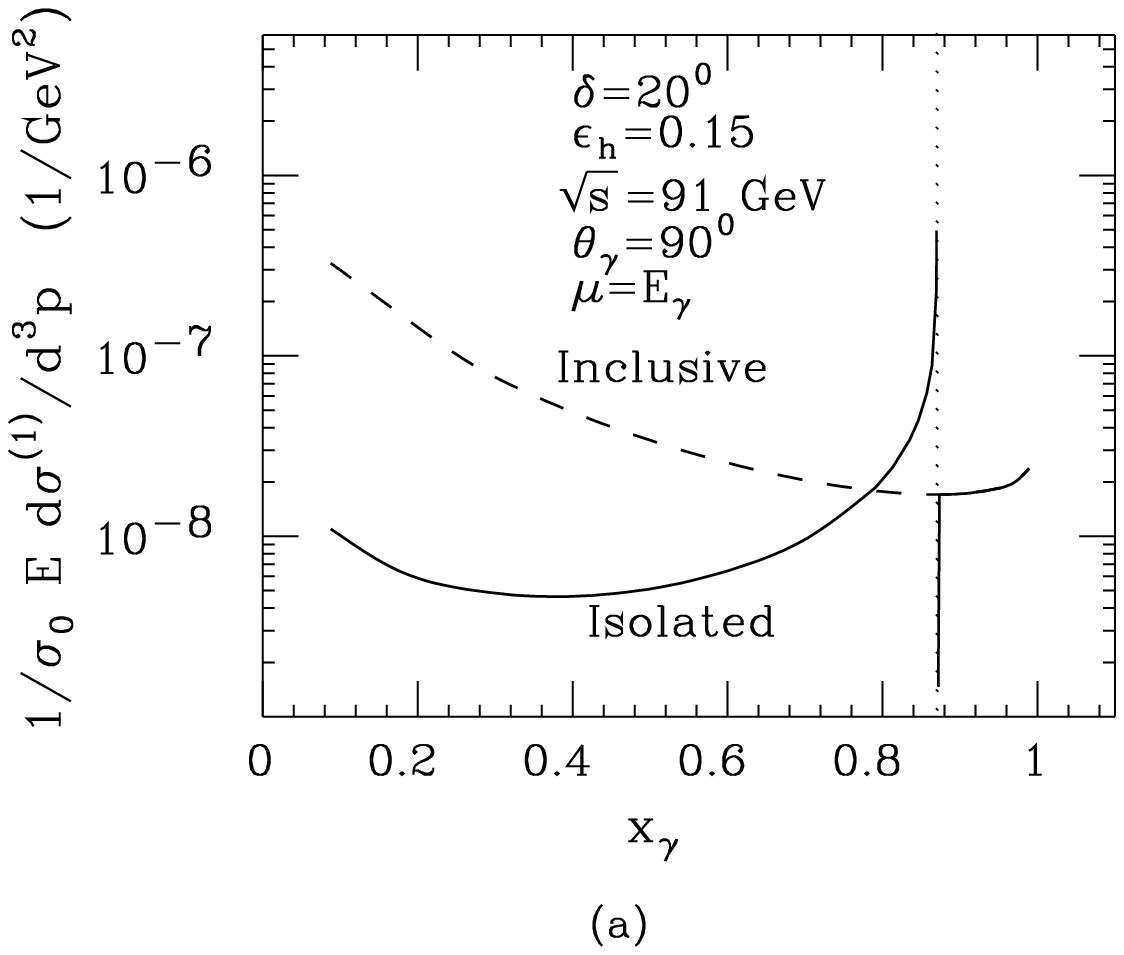}{\hskip -1.0cm}
\epsfxsize9.5cm\epsffile{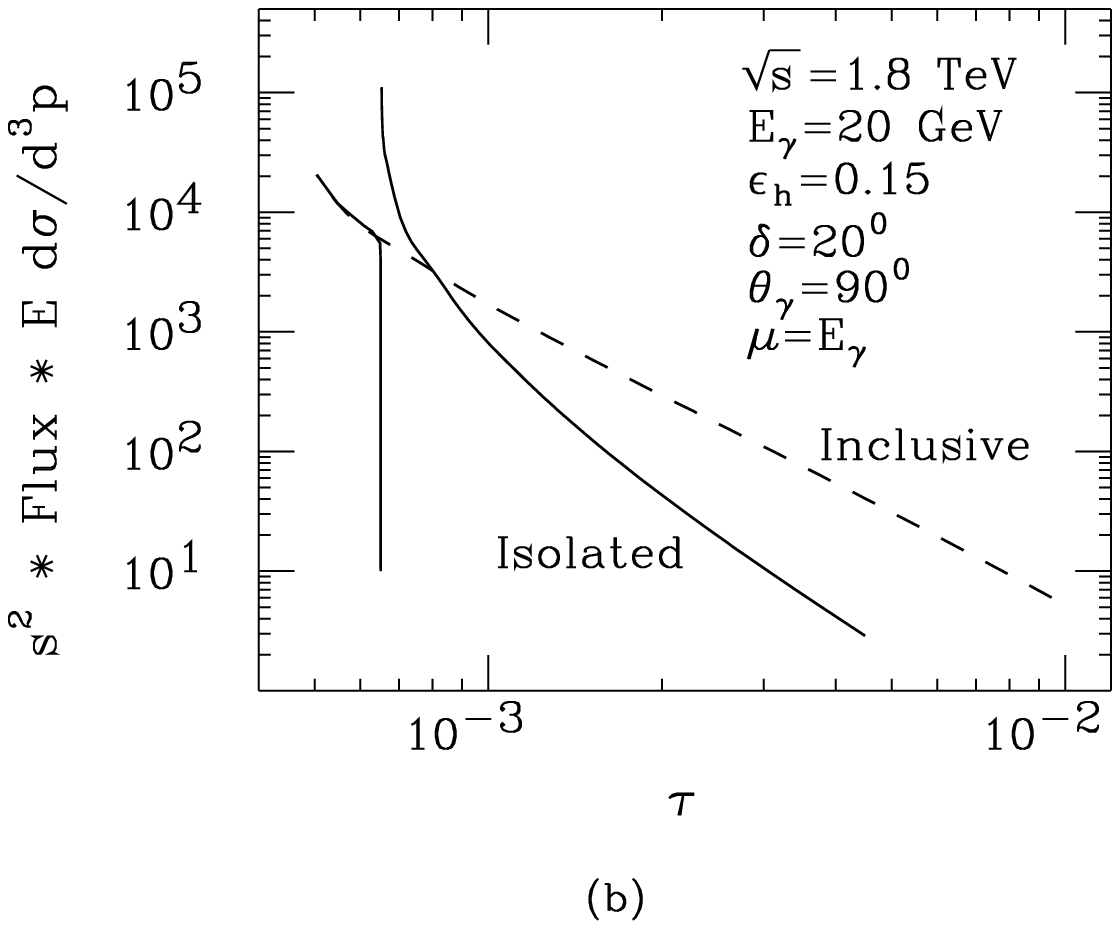}}
\caption{ One-loop quark fragmentation contributions to the isolated 
and inclusive cross sections (a) as a function of 
$x_\gamma = 2 E_\gamma / \protect\sqrt{s}$ in 
$e^+e^- \rightarrow \gamma X$ at $\protect\sqrt{s} = 91$~GeV, and (b) 
as a function of $\tau$ in $p \bar{p} \rightarrow \gamma X$
at $\protect\sqrt{s} =1.8$~TeV.}
\end{figure}

\subsection{$x_\gamma < 1/(1+\epsilon_h)$}
When $x_\gamma < 1/(1+\epsilon_h)$, subprocesses with two-body final
states do not contribute.  This statement follows from energy conservation.  
In a two-body final state, $E_\gamma + E_{\rm{frag}} = \protect\sqrt{s}/2$, 
and $E_{\rm hadrons}^{\rm cone} = E_{\rm frag} = 
(1 - x_{\gamma})\protect\sqrt{s}/2$.  Isolation requires 
$E_{\rm hadrons}^{\rm cone} \le \epsilon_h E_\gamma =  
\epsilon_hx_{\gamma}\protect\sqrt{s}/2$.  Correspondingly, the two-body final 
state processes contribute only for $x_{\gamma} \ge x_{crit}$.  In 
particular, since two-body final states are absent, there is no contribution 
from one-loop virtual gluon exchange diagrams.  There is a contribution from 
the three-body final state real gluon emission diagrams.  At the partonic 
level, these yield a well-known infrared pole singularity having the form 
$1/(1-x_q)$ as $x_q=x_\gamma/z \rightarrow 1$.  Because the 
one-loop virtual gluon exchange diagrams do not contribute, the 
infrared pole singularity {\it remains uncanceled} in 
$\hat{\sigma}^{iso}_{e^+e^-\rightarrow qX}$.  

If conventional factorization were valid, the fragmentation contributions to 
the physical cross section would be expressed in the factorized form
\begin{equation}
E_\gamma \frac{d\sigma^{iso}_{e^+e^- \rightarrow \gamma X}}{d^3\ell}
= \sum_c\  
    \int_{{\rm max}\left[x_\gamma,\frac{1}{1+\epsilon_h}\right]}^1 
         \frac{dz}{z}\
    E_c \frac{d\hat{\sigma}^{iso}_{e^+e^- \rightarrow cX}}{d^3 p_c} 
    \left(x_c=\frac{x_\gamma}{z}\right) 
    \frac{D_{c \rightarrow \gamma}(z,\delta)}{z}\ ;
\label{e1}
\end{equation}
$x_c=2E_c/\sqrt{s}$.  The sum extends over $c = q,\bar{q}$ and $g$.  Function 
$D_{c\rightarrow\gamma}(z,\delta)$ is the nonperturbative function that 
describes fragmentation of parton ``$c$'' into a photon; its 
evolution is governed by the Altarelli-Parisi equations.  The lower limit of 
integration results from the isolation requirement with the assumption that all
fragmentation energy is in the isolation cone~\cite{BGQ2}.  After 
convolution with $D_{q\rightarrow \gamma}(z)$, the inverse power infrared 
sensitivity of $\hat{\sigma}^{iso}_{e^+e^-\rightarrow qX}$, 
at the partonic level, yields a logarithmic divergence in the 
physical cross section $\sigma^{iso}_{e^+e^- \rightarrow \gamma X}$ 
proportional to $\ell n(1/x_\gamma -(1+\epsilon_h))$.  
As shown in Fig.~2(a), this 
means that the isolated cross section would become larger than the inclusive 
cross section in the vicinity of $x_\gamma \rightarrow 1/(1+\epsilon_h)$, a
result that is certainly not physical.  This infrared sensitivity in 
$\hat{\sigma}^{iso}_{e^+e^-\rightarrow qX}$ signals a clear breakdown of
conventional perturbative factorization. 

Our claim that conventional perturbative factorization breaks down 
has been contested~\cite{afgkp}.  However, a careful reading of that paper 
shows that, after repeating our derivation, the authors come to the same 
conclusions as we do.  To quote from the concluding 
paragraph~\cite{afgkp}, ``the [infrared] logarithms .. become large in the 
neighborhood of $x_\gamma \sim x_{crit}$ ..."; ``one has to study whether 
these [infrared] logarithms can be factored since they reflect long-distance 
effects"; and ``they destroy the relevance of the perturbative expansion at 
least in the neighborhood of $x_{crit}$". 

\subsection{$x_\gamma = 1/(1+\epsilon_h)$}
When $x_\gamma = 1/(1+\epsilon_h)$, it is possible to have 
$x_q=x_\gamma/z=1$.  The one-loop virtual gluon exchange diagrams contribute 
fully in this case,  with contributions proportional to $\delta(1-x_q)$.  
However, isolation constraints limit the phase space accessible 
to gluon emission in the real subprocess, $e^+e^-\rightarrow q\bar{q}g$.  
Consequently, the infrared divergences of the real and virtual
contributions do not cancel completely in the isolated case, unlike
the inclusive case.  Working in $n=4-2\epsilon$ dimensions, after 
adding the real and virtual contributions, we find~\cite{BGQ2}
\begin{equation}
E_q \frac{d\hat{\sigma}^{iso(real+virtual)}_{e^+e^-\rightarrow q X}}{d^3p_q} 
\sim
    \left\{\frac{1}{\epsilon^2}+\frac{1}{\epsilon} 
          \left(\frac{3}{2}-\ell n\frac{\delta^2}{4}\right)
         \right\}\, \delta(1-x_q)\ 
  + \mbox{finite terms} \ .
\label{e2}
\end{equation}
The presence of the uncanceled $1/\epsilon$ and $1/\epsilon^2$ terms means 
that the regulator $\epsilon$ cannot be set to 0.  Therefore, 
at $x_q=1$, corresponding to $x_\gamma = 1/(1+\epsilon_h)$, the partonic
term for quark fragmentation is infrared
divergent, and the perturbative calculation is not well-defined.
Conventional perturbative factorization again breaks down.

In Fig.~2(a) we present the perturbatively computed one-loop quark 
fragmentation contributions to the physical cross sections for inclusive and 
isolated prompt photon production $e^+e^- \rightarrow \gamma + X$  
at $\protect\sqrt{s} = 91$~GeV.

\subsection{Relevance for Experiment}
The logarithmic divergence in the physical cross section 
$d\sigma^{iso}/dx_\gamma$ 
for $x_\gamma < x_{crit}$ is an integrable singularity.  Data are presented 
in bins of finite width, related to the experimental resolution.  If the 
perturbative divergence spans a very narrow region in $x_\gamma$, one that is 
smaller than the typical bin width, then the theoretical cross section will 
still be useful.  For the situation examined in this paper, the perturbatively 
calculated isolated cross section exceeds the inclusive cross section over a  
region in $x_\gamma$ that is not narrow.  As 
indicated in Fig.~2(a), the region extends over an interval of about 4 GeV in 
$E_{\gamma}$, larger than typical bin widths at LEP and SLC.  This means that 
the perturbatively computed $d\sigma^{iso}/dx_\gamma$ cannot be accepted at 
face value.  Conventional perturbative QCD leaves strong infrared sensitivity 
in the partonic hard-part at next-to-leading order, and we must look beyond 
fixed-order perturbation theory in order to derive an expression that makes 
physical sense.  

\subsection{Origin and Possible Cure of the infrared Sensitivity}
For $x_{\gamma} < x_{crit}$, the infrared sensitivity of the form 
$\frac{1}{1-x_q}\ln(\frac{1}{1-x_q})$ comes from soft real gluon 
emission.  An analogous case is the transverse momentum ($q_T$) distribution 
for the production of the intermediate vector bosons $W$ and $Z$ or of 
massive lepton-pairs (the Drell-Yan process).  At ${\cal O}(\alpha_s)$, the 
subprocess $q + \bar{q} \rightarrow \gamma^* + g$ provides a singular 
distribution of the form $\alpha_s/q_T^2$.  Likewise, the thrust distribution 
in $e^+e^-$ annihilation, $d\sigma/dT$, is singular in the limit 
$T \rightarrow 1$.  Fixed-order calculations do not work as 
$q_T \rightarrow 0$ or as $T \rightarrow 1$, and resummation is 
invoked~\cite{css2,thrust}.  For isolated photon production, as discussed in 
this paper, a similar problem is encountered as $x_q \rightarrow 1$, or, 
equivalently, as $x_\gamma  \rightarrow \frac{1}{1 + \epsilon_h}$.  All-orders 
resummation of soft gluon radiation is therefore also suggested for isolated 
photon production.  Owing to the limited phase space available for gluon 
radiation, a Sudakov suppression of the infrared divergence might be expected.  

While there are similarities, there are also significant differences between 
the isolated photon case and the $q_T$ distribution and thrust examples cited 
above.  In the isolated photon case, the point of infrared divergence occurs 
within the physical region at a location determined by the experimenters' 
choice of $\epsilon_h$, not at a fixed point at the edge of phase space (i.e., 
$q_T =$ 0, and $T =$ 1).  More importantly, in the Drell-Yan and thrust 
examples, the infrared divergence cancels exactly between the real emission 
and gluon loop diagrams.  Absent in these cases are uncanceled 
$1/\epsilon^2$ and $1/\epsilon$ poles that arise in the isolated photon case 
at $x_\gamma = x_{crit}$ from the restricted phase space for real gluon 
emission, c.f., Eq.~(\ref{e2}).  In the Drell-Yan and thrust cases, the 
divergence is regulated, and soft-gluon resummation can be done in Fourier 
transformed impact parameter space.  The uncanceled poles of the isolated 
photon case would be tantamount to an extra unbalanced $\delta(q_T^2)$ piece 
in the Drell-Yan case.  For isolated photon production, resummation must be 
done at the partonic level, in the $x_q$ distribution, before the convolution 
is performed with the $q \rightarrow \gamma X$ fragmentation function.  A 
resummation procedure must be devised to handle the unbalanced 
$\delta(1 - x_q)$ problem of Eq.~(\ref{e2}).  

The presence of infrared sensitivity is a tip-off that non-perturbative effects 
are present and must be addressed.  For example, in the example of the 
$q_T$ distribution in Drell-Yan case, non-perturbative functions $g_i$ are 
introduced in the implementation of resummation in impact parameter 
space~\cite{css2}.  These unknown functions determine the behavior of the 
differential cross section at modest values of $q_T$.  The critical point 
$x_{crit}$ of the isolated photon case, defined in 
Eq.~(\ref{ec}), is arbitrary since $\epsilon_h$ is chosen experimentally.  If 
$\epsilon_h$ is very small, $x_{crit} \rightarrow 1$; there will be only 
a very narrow region over which resummation will matter, and fixed-order 
perturbation theory will be adequate.  On the other hand, if we are interested 
in extracting fragmentation functions from the data, $\epsilon_h$ must not be 
too small.  There will then be a relatively large region in which 
non-perturbative functions analogous to $g_i$ will play a significant role in 
fits to data.  Their presence 
is a source of uncertainty for the extraction of quark-to-photon fragmentation 
functions in next-to-leading order.  After resummation, instead of the 
divergence apparent in the solid curve in Fig.~2(a), the predicted isolated 
cross section below $x_\gamma = x_{crit}$ will remain bounded from above by 
the inclusive prediction, and it will join smoothly at $x_{crit}$ to the form 
it takes in Fig.~2(a) above $x_{crit}$.  

\subsection{Recapitulation for $e^+e^-\rightarrow\gamma + X$}
To recapitulate, in $e^+e^-\rightarrow\gamma + X$, the next-to-leading order 
partonic term associated with the quark fragmentation contribution 
is infrared sensitive when $x_\gamma \leq 1/(1+\epsilon_h)$.  
Conventional perturbative factorization of the cross section for isolated 
photon production in $e^+e^-$ annihilation breaks down in the neighborhood of 
$x_\gamma = 1/(1+\epsilon_h)$.  The isolated cross section, as usually 
defined, is not an infrared safe observable and cannot be calculated reliably 
in conventional fixed-order perturbative QCD at, and in the immediate region 
below, $x_\gamma=1/(1+\epsilon_h)$.  All-orders resummation offers a possible 
theoretical resolution, but this situation differs in important respects from 
other examples of successful application of resummation techniques.  

\section{Hadron Collider Experiments}
In hadron collisions, $A + B \rightarrow \gamma X$, we are interested in the 
production of isolated prompt photons as a function of the photon's 
transverse momentum, $p_T$.  At next-to-leading order in QCD, one must
include fragmentation at next-to-leading order.  At this order, difficulties  
analogous to those in $e^+e^-$ annihilation are encountered also in 
the hadronic case.  To illustrate the problem~\cite{BGQ2},
we consider the contribution from a quark-antiquark subprocess
in which the flavors of the initial and final quarks differ:  $q' +
\bar{q}' \rightarrow q + \bar{q} + g$, where $q$ fragments to a 
$\gamma$.  We keep only final state gluon radiation so that the results of 
our investigations of $e^+e^-$ annihilation can be exploited directly.  We 
specialize to rapidity $y_{\gamma} =$ 0 and take equal values for the incident 
parton momentum fractions, $x_a = x_b = x = \sqrt{\tau}$.  
In the translation to the hadronic case, the variable $x_\gamma$ 
becomes $\hat{x}_T$ where $\hat{x}_T = 2 p_T/\sqrt{\hat{s}} \sim
x_T/x$ with $x_T=2p_T/\sqrt{s}$.  The critical point at which infrared 
divergence is manifest in the $e^+e^-$ case tends to occur at relatively large 
values of $x_{\gamma}$.  Owing to the correspondence $x_{\gamma} \sim x_T/x$, 
the critical point in the hadron case occurs at small values of the parton 
momentum fraction $x$ where its effects are enhanced by the parton densities.

The special one-loop quark fragmentation contribution
to the observed cross section takes the form 
\begin{equation}
E_\gamma \frac{d\sigma_{AB \rightarrow \gamma X}}{d^3\ell} 
\sim  
\int_{x_T^2}^1 d\tau\ \Phi_{q'\bar{q}'}(\tau)\ 
E_\gamma \frac{d\sigma_{q'\bar{q}'\rightarrow \gamma
X}}{d^3\ell}(\tau) \quad + \quad \mbox{other subprocesses} \ . 
\label{e7st}
\end{equation}
In Fig.~2(b), we show the {\it integrand} in Eq.~(\ref{e7st})
obtained after convolution with the parton flux $\Phi(\tau)$.  We compare 
the integrands for the isolated and the inclusive cases, and we observe again 
that infrared sensitivity at fixed-order leads to the unphysical result that 
the isolated integrand exceeds the inclusive integrand.  It is evident that 
the convolution with the parton flux substantially enhances the influence of 
the region of infrared sensitivity.  

The integrand of Eq.~(\ref{e7st}) is not a physical observable.  Instead, the 
contribution to the hadronic cross section is the area under the curve in 
Fig.~2(b) from $x_T^2$ to 1. The divergences above and below the point
$\hat{x}_T = 1/(1 + \epsilon_h)$ [or
$\sqrt{\tau}=x_T(1+\epsilon_h)$] are integrable logarithmic
divergences, and thus they yield a finite contribution if an
integral is done over all $\tau$.  Indeed, after integration, the calculated 
isolated cross section may well turn out less than its inclusive counterpart 
since the unwarranted extra positive contribution associated with the 
logarithmic divergence in the region of small $\tau$ may be more than 
compensated by the area between the inclusive and the isolated curves at 
large $\tau$.  The issue is one of reliability.  We stress that the 
perturbatively calculated one-loop partonic cross section
$E_q\, d\hat{\sigma}^{iso}_{q'\bar{q}'\rightarrow q X}/d^3p_q$, has an 
inverse-power divergence as $x_q\rightarrow 1$ and has uncanceled 
$1/\epsilon^2$ and $1/\epsilon$ poles in dimensional 
regularization~\cite{BGQ2}.  This pole divergence for 
$\hat{x}_T < 1/(1 + \epsilon_h)$ becomes a logarithmic divergence in 
$E_\gamma d\sigma_{q'\bar{q}'\rightarrow \gamma X}/d^3\ell$ after the 
convolution with a long-distance $q \rightarrow \gamma X$ fragmentation 
function.  Although the logarithmic divergence near 
$\sqrt{\tau}=x_T(1+\epsilon_h)$ is integrable, the isolated integrand should 
never exceed the inclusive integrand.  It is not correct to accept at face 
value a prediction for an 
isolated cross section obtained from a perturbatively 
calculated integrand whose value exceeds that appropriate for the inclusive 
case (even if, after integration over $\tau$, the resulting isolated cross 
section is smaller than the inclusive).  

In the calculated isolated cross section, how large is the uncertainty 
associated with the infrared sensitivity of the next-to-leading order quark 
fragmentation terms?   Referring to Fig.~2(b), we define the overestimate 
of the isolated cross section to be the area under the solid curve but 
above the dashed curve.  We define the maximum isolated cross section to be 
the area under the lower of the solid and dashed curves.  Doing so, we find 
that the next-to-leading order quark-fragmentation contribution to the 
physical isolated cross section is overestimated by 54\%.  While this 
overestimate is certainly large, one should bear in mind that the tree-level 
fragmentation contribution is not affected by the infrared uncertainty and 
that the tree-level term is larger than the one-loop quark fragmentation term 
that is of concern.  Thus, the uncertainty in the overall fragmentation 
contribution will be typically only some fraction of 54\%.  Second, even 
though fragmentation may account for half or more of the inclusive prompt 
photon yield at relatively small values of $p_T$, the fragmentation fraction is 
substantially reduced by isolation.  We suggest therefore, that the net 
uncertainty in the physical isolated cross section associated with the 
next-to-leading order fragmentation terms may be at the 10\%\ level.  

\section{Summary}
The results in both the $e^+e^-$ and hadronic cases challenge us to find 
a modified factorization scheme and/or to devise more appropriate infrared
safe observables.

\section*{Acknowledgments}
Work at Argonne National Laboratory is supported in part by the U. S. 
Department of Energy, Division of High Energy Physics, Contract 
No. W-31-109-ENG-38.



\section*{References}
\vspace{-0.2cm}
\begin{thebibliography}{99}


\bibitem{css1}
 J.~C.~Collins, D.~E.~Soper, and G.~Sterman, in {\it Perturbative
 Quantum Chromodynamics}, ed. A.~H.~Mueller (World Scientific, Singapore, 
 1989).

\bibitem{BGQ1}
 E.~L.~Berger, X.~F.~Guo, and J.~W.~Qiu, \Journal{\PRD} {53} {1124} {1996}.

\bibitem{BGQ2} 
 E.~L.~Berger, X.~F.~Guo, and J.~W.~Qiu, \Journal{\PRL}{76}{2234}{1996}; 
 \Journal{\PRD} {54} {5470} {1996}.

\bibitem{theory}
 P.~Aurenche {\it et al}, \Journal{\NPB} {399} {34} {1993} and references 
 therein; 
 H.~Baer, J.~Ohnemus, and J.~F.~Owens, \Journal{\PRD} {42} {61} {1990}; 
 E.~L.~Berger and J.~W.~Qiu, \Journal{\PLB} {248} {371} {1990} and 
 \Journal{\PRD} {44} {2002} {1991}; 
 L.~Gordon and W.~Vogelsang, \Journal{\PRD} {50} {1901} {1994}; 
 L. Gordon, Argonne report ANL-HEP-PR-96-60 (hep-ph/9611391), November, 1996.  

\bibitem{exptal1}
 ALEPH Collaboration, D. Buskulic {\it et al}, \Journal{\ZPC} {57} {17} {1993} 
 and references therein; 
 DELPHI Collaboration, P. Abreu {\it et al}, \Journal{\ZPC} {69} {1} {1995} and 
 references therein;
 L3 Collaboration, M. Acciarri {\it et al}, \Journal{\PLB} {388} {409} {1996} 
 and references therein; 
 OPAL Collaboration, G. Alexander {\it et al}, \Journal{\ZPC} {71} {1} {1996} 
 and references therein.

\bibitem{early} 
K.~Koller, T.~F.~Walsh, and P.~M.~Zerwas, \Journal{\ZPC} {2} {197} {1979};
E.~Laermann, T.~F.~Walsh, I.~Schmitt, and P.~M.~Zerwas, \Journal{\NPB} {207} 
{205} {1982};  E.~W.~N.~Glover and A.~G.~Morgan, \Journal{\ZPC} {62} {311} 
{1994}; G.~Kramer and B.~Lampe, \Journal{\ZPC} {34} {497} {1987} and  
\Journal{\PLB} {269} {401} {1991};
 G.~Kramer and H.~Spiesberger, {\it Proceedings of Annecy Workshop on Photon 
 Radiation from Quarks}, Annecy, France, December 1991,   
 CERN report CERN 92-04, ed. S. Cartwright, pp. 26-40;
P.~M\"attig, H.~Spiesberger, and W.~Zeuner, \Journal{\ZPC} {60} {613} {1993}.

\bibitem{bgq3}
E.~L.~Berger, X.~F.~Guo, and J.~W.~Qiu, in {\it The Fermilab Meeting}, 
Proceedings of the 7th Meeting of the Division of Particles and Fields of the 
American Physical Society, Fermilab, November, 1992; ed. C.~H.~Albright, 
P.~H.~Kasper, R.~Raja, and J.~Yoh (World Scientific, Singapore, 1993), 
pp. 957-960.

\bibitem{exptal2}
 ALEPH Collaboration, D. Buskulic {\it et al}, \Journal{\ZPC} {69} {365} 
{1996}; OPAL Collaboration, K. Ackerstaff {\it et al}, CERN report 
CERN-PPE/97-086 (July, 1997), submitted to {\it Z.~Phys.} C.  

\bibitem{afgkp}
 P.~Aurenche, M.~Fontannaz, J.~Ph.~Guillet, A.~Kotikov, and E.~Pilon, 
 \Journal{\PRD} {55} {1124} {1997}.

\bibitem{css2}
 J.~C.~Collins, D.~E.~Soper, and G.~Sterman, \Journal{\NPB} {250} {199} {1985}.
 For a recent treatment and list of references, see R.~K.~Ellis, D.~A.~Ross, 
 and S.~Veseli, Fermilab report FERMILAB-PUB-97/082-T (hep-ph/9704239), April, 
 1997.  
\bibitem{thrust}
 M.~Greco and Y.~Srivastava, \Journal{\PRD} {23} {2791} {1981}; S.~Catani, 
G.~Turnock, B.~R.~Webber, and L. Trentadue, 
\Journal{\PLB} {263} {491} {1991}; G.~Sterman, in {\it QCD and 
Beyond}, Theoretical Advanced Study Institute in Elementary Particle 
Physics (TASI'95), June, 1995, ed. D.~E.~Soper (World Scientific, Singapore, 
1995).  

\end{thebibliography}
\end{document}